\documentclass[12pt]{article}

\usepackage[english]{babel}
\usepackage{amsfonts,amsmath,amssymb,amscd}
\usepackage{bbm}

\input xy
\xyoption{all}

\makeatletter\renewcommand{\section}%
  {\@startsection {section}{1}{\z@}{-3.5ex plus -1ex minus-.2ex}%
  {2.3ex plus .2ex}{\bf }}
\makeatletter\renewcommand{\subsection}%
  {\@startsection{subsection}{2}{\z@}{-3.25ex plus -1ex minus -.2ex}%
  {1.5ex plus .2ex}{\it }}
\makeatletter\renewcommand{\paragraph}%
  {\@startsection{paragraph}{3}{\z@}{.25ex \@plus1ex \@minus.2ex}{-.5em}{\bf }}

\numberwithin{equation}{section}

\renewenvironment{thebibliography}[1]
     { \baselineskip=16pt plus 2pt minus 1pt 
      \section*{\centerline{\refname}
        \@mkboth{\MakeUppercase\refname}{\MakeUppercase\refname}}%
     \footnotesize\list{\@biblabel{\@arabic\c@enumiv}}%
           {\settowidth\labelwidth{\@biblabel{#1}}%
            \leftmargin\labelwidth
            \advance\leftmargin\labelsep
            \@openbib@code
            \usecounter{enumiv}%
            \let\p@enumiv\@empty
            \renewcommand\theenumiv{\@arabic\c@enumiv}}%
      \sloppy
      \clubpenalty4000
      \@clubpenalty \clubpenalty
      \widowpenalty4000%
      \sfcode`\.\@m}

\makeatother

\newtheorem{Cor}{Corollary}
\newtheorem{Thm}{Theorem}

\newcommand{\CA}{\mathcal{A}} 
\newcommand{\CE}{\mathcal{E}} 
\newcommand{\CF}{\mathcal{F}}
\newcommand{\CL}{\mathcal{L}}
\newcommand{\CN}{\mathcal{N}}
\newcommand{\CO}{\mathcal{O}}  
\newcommand{\CP}{\mathcal{P}} 
\newcommand{\CU}{\mathcal{U}} 

\newcommand{\IC}{\mathbbm{C}}    
\newcommand{\IZ}{\mathbbm{Z}}  
\newcommand{\pbar}{\bar{\partial}}   
\renewcommand{\a}{{\alpha}}
\newcommand{\ad}{{\dot{\alpha}}}  
\renewcommand{\b}{{\beta}}
\newcommand{\bd}{{\dot\beta}} 
\newcommand{\g}{{\gamma}}
\newcommand{\gd}{{\dot\gamma}} 
\renewcommand{\o}{{\omega}} 
\renewcommand{\d}{{\delta}}
\renewcommand{\l}{\lambda}  
\newcommand{\hl}{{\hat{\lambda}}}

\newcommand{\eqn}[2]{\begin{equation}\label{#1}#2\end{equation}}
\newcommand{\eqna}[2]{\begin{equation}\label{#1}
                      \begin{aligned}#2\end{aligned}\end{equation}}

\begin{document}
\begin{titlepage}
\setcounter{page}{0}
\begin{flushright}
  hep-th/0511230\\
  ITP--UH--24/05
\end{flushright}
\vspace*{3cm}
\begin{center}
{\LARGE \bf Twistors and Aspects of Integrability\\[4pt] 
            of self-dual SYM Theory} 
\vskip 1.5cm
{\large Martin Wolf} \footnote{\ E-mail: {\ttfamily 
 wolf@itp.uni-hannover.de} }
\vskip .8cm
{\em Institut f\"ur Theoretische Physik\\
     Universit\"at Hannover\\
     Appelstra{\ss}e 2, 30167 Hannover, Germany}
\vskip 1cm
\end{center}
\begin{center}
{\bf Abstract}
\end{center}
With the help of the Penrose-Ward transform, which relates certain 
holomorphic vector bundles over the supertwistor space to the equations 
of motion of self-dual SYM theory in four dimensions, we construct 
hidden infinite-dimensional symmetries of the theory. 
We also present a new and shorter proof
(cf. hep-th/0412163) of the relation between certain deformation algebras and  
hidden symmetry algebras. This article is based on a talk given by
the author at the {\it Workshop on Supersymmetries and Quantum Symmetries 2005}
at the BLTP in Dubna, Russia.

\begin{quote}
\vskip 5mm
\end{quote}
\vfill
\noindent November, 2005
\end{titlepage}

\newpage
\setcounter{page}{1}

%--------------------------------------------------------------------------
\section{Introduction and conclusions}

By analyzing the linearized \cite{Witten:2003nn} and full \cite{Popov:2004rb}
field equations and by virtue of the Penrose-Ward transform 
\cite{Penrose:1977in}, 
it was shown that there is a one-to-one correspondence 
between the moduli space of holomorphic Chern-Simons theory on supertwistor 
space and of self-dual $\CN=4$ SYM theory in four dimensions.\footnote{For 
reviews of twistor theory, we refer to
\cite{Penrose:ca,Manin:ds,Ward:vs,Mason:1991rf}.} This 
correspondence has then been used for a twistorial construction
of hidden infinite-dimensional symmetry algebras in the self-dual truncation
of SYM theory \cite{Wolf:2004hp}.  Therein, the results known for the purely 
bosonic self-dual YM equations (see, e.g., Refs.  
\cite{Pohlmeyer:1979ya}--\cite{Popov:1998}) have been generalized to the 
supersymmetric setting. Here, we shall briefly report on those results thereby 
also presenting a new and shorter proof of the relation between certain 
deformation algebras on the twistor side and symmetry algebras on the gauge
theory side. For the sake of clarity, the discussion presented below is given
in the complex setting only 
but, of course, it is also possible to implement real
structures (see, e.g., \cite{Popov:2004rb,Wolf:2004hp} for details). \\ 
{\bf Acknowledgments:}
I am grateful to the organizers of the Workshop on Supersymmetries and Quantum 
Symmetries 2005 for inviting me and for the kind hospitality
during the workshop. I would also like to thank A. D. Popov and R. 
Wimmer for commenting on the manuscript. This work was done within the 
framework of the DFG priority program (SPP 1096) in string theory.

\section{Preliminaries}

\subsection{Supertwistor space}

The starting point of our discussion is the complex projective supertwistor
space $\IC P^{3|\CN}$ with homogeneous coordinates $[z^\a,\l_\ad,\eta_i]$
obeying the equivalence relation 
\eqn{}{(z^\a,\l_\ad,\eta_i)\ \sim\ (tz^\a,t\l_\ad,t\eta_i)} 
for any $t\in\IC^*$. Here, the spinorial indices
$\a,\b,\ldots,\ad,\bd,\ldots$ run from $1$ to $2$ and the $R$-symmetry indices
$i,j,\ldots$ from $1$ to $\CN$. In the following, we are interested in the 
open subset $\CP^{3|\CN}:=\IC P^{3|\CN}\setminus\IC P^{1|\CN}$ defined by
$\l_\ad\neq0$. This space can be covered by two patches, say $\CU_+$ and 
$\CU_-$, for which $\l_{\dot{1}}\neq0$ and $\l_{\dot{2}}\neq0$, respectively.
On those patches we have the coordinates 
\eqna{}{z^\a_+\ :=\ \frac{z^\a}{\l_{\dot{1}}},\qquad
       z^3_+\ :=\ \frac{\l_{\dot{2}}}{\l_{\dot{1}}}\ =:\ \l_+
      &\qquad{\rm and}\qquad
      \eta^+_i\ :=\ \frac{\eta_i}{\l_{\dot{1}}}\qquad{\rm on}\qquad\CU_+,\\
      z^\a_-\ :=\ \frac{z^\a}{\l_{\dot{2}}},\qquad
       z^3_-\ :=\ \frac{\l_{\dot{1}}}{\l_{\dot{2}}}\ =:\ \l_-
      &\qquad{\rm and}\qquad
      \eta^-_i\ :=\ \frac{\eta_i}{\l_{\dot{2}}}\qquad{\rm on}\qquad\CU_-,
}
which are related by
\eqn{}{ z^\a_+\ =\ \frac{1}{z^3_-}z^\a_-,\qquad z^3_+\ =\ \frac{1}{z^3_-}
   \qquad{\rm and}\qquad\eta^+_i\ =\ \frac{1}{z^3_-}\eta^-_i}
on $\CU_+\cap\CU_-$.
This in particular shows that $\CP^{3|\CN}$, which we simply call
supertwistor space, is a holomorphic fibration over
the Riemann sphere $\IC P^1$, 
\eqn{FIB}{\CP^{3|\CN}\ =\ \CO(1)\otimes\IC^2\oplus\Pi\CO(1)\otimes\IC^\CN
   \ \to\ \IC P^1.}

From this definition it is clear that global holomorphic sections of the 
fibration \eqref{FIB} 
are degree one polynomials. In a given trivialization, they
are locally of the form
\eqn{HS1}{z^\a_\pm\ =\ x^{\a\ad}\l_\ad^\pm\qquad{\rm and}\qquad
       \eta_i^\pm\ =\ \eta^\ad_i\l_\ad^\pm}
and parametrized by the moduli $(x^{\a\ad},\eta^\ad_i)\in\IC^{4|2\CN}$. 
Here, we also introduced the common abbreviations 
\eqn{}{(\l_\ad^+)\ :=\ \binom{1}{\l_+}\qquad{\rm and}\qquad
       (\l_\ad^-)\ :=\ \binom{\l_-}{1}.}
Therefore, $\CP^{3|\CN}$ naturally fits into the following 
double fibration
\eqna{DF1}{
 \begin{picture}(50,40)
  \put(0.0,0.0){\makebox(0,0)[c]{$\CP^{3|\CN}$}}
  \put(64.0,0.0){\makebox(0,0)[c]{$\IC^{4|2\CN}$}}
  \put(34.0,33.0){\makebox(0,0)[c]{$\CF^{5|2\CN}$}}
  \put(7.0,18.0){\makebox(0,0)[c]{$\pi_2$}}
  \put(55.0,18.0){\makebox(0,0)[c]{$\pi_1$}}
  \put(25.0,25.0){\vector(-1,-1){18}}
  \put(37.0,25.0){\vector(1,-1){18}}
 \end{picture}
}
where $\CF^{5|2\CN}\cong\IC^{4|2\CN}\times\IC P^1$ is called the 
correspondence space. The (holomorphic) projections are given according to
\eqna{}{ \pi_1\,:\,(x^{\a\ad},\l_\pm,\eta^\ad_i)\ 
          &\mapsto\ (x^{\a\ad},\eta^\ad_i),\\
        \pi_2\,:\,(x^{\a\ad},\l_\pm,\eta^\ad_i)\ 
          &\mapsto\ (z^\a_\pm=x^{\a\ad}\l^\pm_\ad,z^3_\pm=\l_\pm,
                     \eta^\pm_i=\eta^\ad_i\l^\pm_\ad).       
}

Next let us take a closer look at the relations \eqref{HS1}. Fixing a point 
$(z^\a_\pm,\l_\pm,\eta^\pm_i)$ in supertwistor space and solving \eqref{HS1}
for $(x^{\a\ad},\eta^\ad_i)$, one determines an isotropic (null) plane
$\IC^{2|\CN}$ in $\IC^{4|2\CN}$. On the other hand, a fixed point
$(x^{\a\ad},\eta^\ad_i)\in\IC^{4|2\CN}$ gives a holomorphic embedding
of the Riemann sphere into supertwistor space. Thus, we have
\begin{center} 
 \begin{tabular}{cccc}
   {\rm (i)} & a point $p\in\CP^{3|\CN}$ & $\quad\longleftrightarrow\quad$ & 
        an isotropic plane $\IC_p^{2|\CN}\hookrightarrow\IC^{4|2\CN}$,\\
   {\rm (ii)} & $\IC P^1_{x,\eta}\hookrightarrow\CP^{3|\CN}$&$\quad
                \longleftrightarrow\quad$ & 
                 a point $(x,\eta)\in\IC^{4|2\CN}$.
 \end{tabular}
\end{center}

\subsection{Holomorphy and self-dual SYM theory in the twistor approach}

In order to study super gauge theory, some additional data on the manifolds 
appearing in the double fibration \eqref{DF1} is required. Let us consider a 
rank $n$ holomorphic vector bundle $\CE\to\CP^{3|\CN}$ which is characterized
by the transition function $f=\{f_{+-}\}$ and its pull-back
$\pi_2^*\CE$ to the supermanifold $\CF^{5|2\CN}$. For notational
reasons, we denote the pulled-back transition function by the same letter $f$.
By definition of a pull-back, the transition function $f$ is constant along
the fibers of $\pi_2\,:\CF^{5|2\CN}\to\CP^{3|\CN}$. Therefore, 
it is annihilated by the vector fields 
\eqn{HVF}{D^\pm_\a\ :=\ \l^\ad_\pm\partial_{\a\ad}\qquad{\rm and}\qquad
       D^i_\pm\ :=\ \l^\ad_\pm \partial^i_\ad,}
where $\partial_{\a\ad}=\partial/\partial x^{\a\ad}$ and 
$\partial^i_\ad=\partial/\partial\eta^\ad_i$. Spinorial indices are raised and 
lowered via the $\epsilon$-tensors, $\epsilon^{12}=\epsilon^{{\dot1}{\dot2}}=
-\epsilon_{12}=-\epsilon_{{\dot1}{\dot2}}=1$, together with the normalizations 
$\epsilon_{\a\b}\epsilon^{\b\g}=\d^\g_\a$ and 
$\epsilon_{\ad\bd}\epsilon^{\bd\gd}=\d^\gd_\ad$. Let $\pbar_\CP$ and 
$\pbar_\CF$ be the anti-holomorphic parts of the 
exterior derivatives on the supertwistor space and the correspondence 
space, respectively. Then we have $\pi^*_2\pbar_\CP=\pbar_\CF\circ\pi_2^*$, 
and hence, the transition function of $\pi^*_2\CE$ is also annihilated by 
$\pbar_\CF$.

Next we want to assume that the bundle $\CE\to\CP^{3|\CN}$ is holomorphically 
trivial when restricted to any projective line
$\IC P^1_{x,\eta}\hookrightarrow\CP^{3|\CN}$. This condition implies that there
exist some smooth $GL(n,\IC)$-valued functions $\psi=\{\psi_\pm\}$, which 
define a trivialization of $\pi_2^*\CE$, such that $f=\{f_{+-}\}$ can be 
decomposed as
\eqn{SF}{f_{+-}\ =\ \psi^{-1}_+\psi_-}
and 
\eqn{}{\pbar_\CF\psi_\pm\ =\ 0.}
In particular, this formula implies that the $\psi_\pm$ depend holomorphically 
on $\l_\pm$. 
Applying the vector fields \eqref{HVF} to \eqref{SF}, we realize 
by virtue of an extension of Liouville's theorem that
the expressions
\eqn{}{\psi_+ D^+_\a\psi_+^{-1}\ =\ \psi_- D^+_\a
       \psi_-^{-1}\qquad{\rm and}\qquad
       \psi_+ D_+^i\psi_+^{-1}\ =\ \psi_- D_+^i\psi_-^{-1}
}
must be at most linear in $\l_\pm$. Therefore, we may introduce
a Lie-algebra valued one-form $\CA$ such that
\eqna{DefofA}{D^\pm_\a\lrcorner\CA\ &:=\ \CA^\pm_\a\ 
            :=\ \l^\ad_\pm\CA_{\a\ad}\ =\ \psi_\pm D^\pm_\a\psi_\pm^{-1},\\
       D_\pm^i\lrcorner\CA\ &:=\ \CA_\pm^i\ 
            :=\ \l^\ad_\pm\CA_\ad^i\ =\ \psi_\pm D_\pm^i\psi_\pm^{-1},\\
}
and hence
\eqna{LS1}{\l^\ad_\pm(\partial_{\a\ad}+\CA_{\a\ad})\psi_\pm\ &=\ 0,\\
         \l^\ad_\pm(\partial^i_\ad+\CA_\ad^i)\psi_\pm\ &=\ 0}
and $\pbar_\CF\psi_\pm=0$.
The compatibility conditions for the linear system \eqref{LS1} read as
\eqn{CC1}{[\nabla_{\a(\ad},\nabla_{\b\bd)}]\ =\ 0,\quad
          [\nabla^i_{(\ad},\nabla_{\a\bd)}]\ =\ 0\quad{\rm and}\quad
          \{\nabla^i_{(\ad},\nabla^j_{\bd)}\}\ =\ 0,
}
where we have introduced
\eqn{DEFofCD}{\nabla_{\a\ad}\ :=\ \partial_{\a\ad}+\CA_{\a\ad}
              \qquad{\rm and}\qquad
              \nabla^i_\ad\ :=\ \partial^i_\ad+\CA^i_\ad.}
Eqs.  \eqref{CC1} have been known for quite some time and it has been
shown that they are equivalent to the equations of motion of $\CN$-extended 
self-dual SYM theory \cite{Semikhatov:1982ig,Devchand:1996gv} on
four-dimensional space-time.
Note that Eqs. \eqref{LS1} imply that the gauge potentials $\CA_{\a\ad}$ and 
$\CA^i_\ad$ do not change when we perform transformations of the form
$ \psi_\pm \mapsto\psi_\pm h_\pm,$ where the $h=\{h_\pm\}$ are annihilated by 
the vector fields \eqref{HVF} and $\pbar_\CF$. Under such transformations the 
transition function $f=\{f_{+-}\}$ of $\pi_2^*\CE$ transform into a transition 
function $h_+^{-1}f_{+-}h_-$ of a bundle which is said to be equivalent to 
$\pi_2^*\CE$. On the other hand, gauge transformations of the gauge potentials 
are induced by transformations of the form $\psi_\pm\mapsto g^{-1}\psi_\pm$ for
some smooth $\l$-independent $GL(n,\IC)$-valued $g$. Under such transformations
the transition function $f$ is unchanged. In fact, we have 
{\Thm There is a one-to-one correspondence between equivalence classes of 
      holomorphic vector bundles over the supertwistor space which are 
      holomorphically trivial when restricted to any $\IC P^1_{x,\eta}
      \hookrightarrow\CP^{3|\CN}$ and gauge equivalence classes of solutions
      to the equations of motion of $\CN$-extended self-dual SYM theory in 
      four dimensions. In fact, Eqs. \eqref{DefofA} give the Penrose-Ward
      transform, i.e., the relation between fields on supertwistor space and 
      fields on space-time.}

\section{Hidden symmetries}

\subsection{Infinitesimal deformations}

In order to study solutions to the linearized equations of motion
(i.e., symmetries), one
considers small perturbations of the transition functions $f=\{f_{+-}\}$ of a 
holomorphic vector bundle $\CE\to\CP^{3|\CN}$ and its pull-back 
$\pi_2^*\CE\to\CF^{5|2\CN}$, respectively. Note that any infinitesimal 
perturbation 
of $f$ is allowed, as small enough perturbations will, by Kodaira's theorem on 
deformation theory, preserve its trivializability properties on the curves 
$\IC P^1_{x,\eta}\hookrightarrow\CP^{3|\CN}$. This 
 follows directly from $H^1(\IC P^1,\CO)=0$. Thus, we find
\eqn{Deform1}{f_{+-}+\d f_{+-}\ =\ (\psi_++\d\psi_+)^{-1}(\psi_-+\d\psi_-)}
for the deformed transition function of $\pi_2^*\CE$. 
Upon introducing the Lie-algebra valued function 
\eqn{Deform2}{\phi_{+-}\ :=\ \psi_+(\d f_{+-})\psi_-^{-1},}
and linearizing Eq. \eqref{Deform1}, we have to find the splitting
\eqn{PHI1}{\phi_{+-}\ =\ \phi_+-\phi_-.}
Here, the Lie-algebra valued functions $\phi_\pm$ can be extended to 
holomorphic functions in $\l_\pm$ on the respective patches, and which 
eventually yield
\eqn{PHI2}{\d\psi_\pm\ =\ -\phi_\pm\psi_\pm.}
Moreover, we point out that finding such $\phi_\pm$ from $\phi_{+-}$ 
means to solve the infinitesimal variant of the Riemann-Hilbert  problem.
Obviously, the splitting \eqref{PHI1} and hence 
solutions to the Riemann-Hilbert problem are not unique, as we certainly have 
the freedom to consider new ${\tilde\phi}_\pm$ shifted by some function
function $\o$ which is globally defined, i.e., ${\tilde\phi}_\pm=\phi_\pm+\o$.
In fact, such shifts eventually correspond to infinitesimal gauge 
transformations.

Infinitesimal variations of the linear system \eqref{LS1} yield
\eqn{DeformA1}{\d\CA^\pm_\a\ =\ \l^\ad_\pm\d\CA_{\a\ad}\ =\ 
          \l^\ad_\pm\nabla_{\a\ad}\phi_\pm\qquad{\rm and}\qquad
       \d\CA^i_\pm\ =\ \l^\ad_\pm\d\CA^i_\ad\ =\ 
             \l^\ad_\pm\nabla_\ad^i\phi_\pm,}
where the covariant derivatives have been introduced in \eqref{DEFofCD}.
Note that they act adjointly in these equations. The $\l$-expansion of Eqs. 
\eqref{DeformA1} eventually gives the infinitesimal transformation 
$\d\CA_{\a\ad}$ and $\d\CA^i_\ad$, which satisfy by 
construction the linearized equations of motion.
Note that the equivalence relations as defined at the end of 
Sec.~2.2 have an infinitesimal counterpart. Therefore, we altogether have
{\Cor There is a one-to-one correspondence between equivalence classes 
      of deformations of the transition functions of holomorphic vector bundles
      over the supertwistor space which are holomorphically trivial when 
      restricted to any $\IC P^1_{x,\eta}
      \hookrightarrow\CP^{3|\CN}$ and equivalence classes of symmetries 
      of $\CN$-extended self-dual SYM theory in four dimensions.
}\vskip 3mm

\subsection{Hidden symmetry algebras}

Suppose we are given some indexed set $\{\d_a\}$ of infinitesimal
deformations $\d_a f_{+-}$ of the transition function of our holomorphic
vector bundle $\pi^*_2\CE$. Suppose further that the $\d_a$s satisfy 
a deformation algebra of the form 
\eqn{defalg}{[\d_a,\d_b\}f_{+-}\ =\ {f_{ab}}^c\d_cf_{+-},} 
where the ${f_{ab}}^c$s
are generically structure functions and $[\cdot,\cdot\}$ denotes the graded
commutator. Let us assume that the ${f_{ab}}^c$s are constant.
Above we have seen that any such deformation $\d_a f_{+-}$ yields 
a symmetry of $\CN$-extended
self-dual SYM theory. So, given such an algebra, what is the corresponding
symmetry algebra on the gauge theory side? To answer this question, we consider
\eqn{}{[\d_1,\d_2]\ =\ (-)^{p_ap_b}\varepsilon^a\varrho^b[\d_a,\d_b\},}
where $\varepsilon^a$ and $\varrho^b$ are the infinitesimal parameters
of the transformations $\d_1$ and $\d_2$, respectively, and $p_a$ denotes the
Gra{\ss}mann parity of the transformation $\d_a$. Explicitly, we may write
\eqn{COM2}{{[\d_1,\d_2]}\CA^\pm_\a\ =\ \d_1(\CA^\pm_\a+\d_2\CA^\pm_\a)-
           \d_1\CA^\pm_\a-\d_2(\CA^\pm_\a+\d_1\CA^\pm_\a)+\d_2\CA^\pm_\a}
and similarly for $\CA^i_\pm$; cf. also \eqref{DeformA1}. Then one easily 
checks that
\eqn{comofa}{{[\d_1,\d_2]}\CA^\pm_\a\ =\ \l^\ad_\pm
              \nabla_{\a\ad}\Sigma_{12}^\pm,
             \qquad{\rm with}\qquad
            \Sigma_{12}^\pm\ :=\ \d_1\phi^2_\pm-\d_2\phi_\pm^1+
                [\phi^1_\pm,\phi^2_\pm].}
Note that we use the notation $\phi^1_\pm=\varepsilon^a\phi_{\pm a}$ and
similarly for $\phi^2_\pm$.
Next one considers the commutator
\eqn{comoff}{{[\d_1,\d_2]}f_{+-}\ =\ \d_1(f_{+-}+\d_2f_{+-})-\d_1f_{+-}
       -\d_2(f_{+-}+\d_1f_{+-})+\d_2f_{+-}.}
Using the definition \eqref{Deform2} and the resulting splittings \eqref{PHI1}
for the deformations $\d_{1,2}f_{+-}$, one can show after some tedious but 
straightforward algebraic manipulations that the commutator \eqref{comoff} is 
given by
\eqn{result}{{[\d_1,\d_2]f_{+-}}\ =\ \psi_+^{-1}(\Sigma^+_{12}-
              \Sigma^-_{12})\psi_-,}
where $\Sigma^\pm_{12}$ has been introduced \eqref{comofa}. By hypothesis
\eqref{defalg}, it must also be equal to
\eqn{}{{[\d_1,\d_2]f_{+-}}\ =\ \d_3 f_{+-},\qquad{\rm with}\qquad \d_3\ 
              =\ (-)^{p_ap_b}\varepsilon^a\varrho^b {f_{ab}}^c\d_c,}
i.e., 
\eqn{}{{[\d_1,\d_2]f_{+-}}\ =\ \psi_+^{-1}(\phi^3_+-\phi^3_-)\psi_-,}
where $\phi^3_\pm=(-)^{p_ap_b}\varepsilon^a\varrho^b {f_{ab}}^c\phi_{\pm c}$.
By comparing this equation with the result \eqref{result},
we therefore conclude 
\eqn{}{\Sigma^\pm_{12}\ =\ \phi^3_\pm + \o^3\ =\ 
       (-)^{p_ap_b}\varepsilon^a\varrho^b ({f_{ab}}^c\phi_{\pm c}+\o_{ab}),}
since the ${f_{ab}}^c$s are assumed to be constant. Here,
$\o^3$ (respectively, $\o_{ab}$) is some function independent of $\l_\pm$ and
therefore representing  an infinitesimal gauge 
transformation. Combining this result with Eq. \eqref{comofa}, we get the 
following
{\Thm Suppose we are given a deformation algebra of the form
      \eqref{defalg} with constant ${f_{ab}}^c$. Then the corresponding
      symmetry algebra on the gauge theory side has exactly the same
      form modulo possible gauge transformations.}\vskip 2mm
\noindent It should be stressed that this theorem does, however, not give the 
explicit form of the
gauge parameter $\o_{ab}$. In order to compute it, one has to perform the 
splitting procedure explicitly; see \cite{Wolf:2004hp} for details.

\subsection{Examples: Affine extensions of gauge type and superconformal
symmetries}

So far, we have been quite general. Let us now exemplify our discussion.
Let $X_a$ be some generator of the gauge algebra
$\mathfrak{gl}(n,\IC)$ and consider 
\eqn{PRIMEDEF}{\d^m_a f_{+-}\ :=\ \l_+^m[X_a,f_{+-}],
      \qquad{\rm for}\qquad m\in\IZ.}
For $m=0$, the transformations of the components of the gauge potential
are given by
\eqn{}{\d^0_a\CA_{\a\ad}\ =\ [X_a,\CA_{\a\ad}]\qquad{\rm and}\qquad
       \d^0_a\CA^i_\ad\ =\ [X_a,\CA^i_\ad].}
Thus, they represent a gauge type transformation
with constant gauge parameter (a global symmetry transformation). 
The corresponding deformation algebra is easily computed to be
\eqn{}{[\d^m_a,\d^n_b]f_{+-}\ =\ {f_{ab}}^c\d^{m+n}_cf_{+-},}
where the ${f_{ab}}^c$s are the structure constants of $\mathfrak{gl}(n,\IC)$,
i.e., we get a centerless Kac-Moody algebra. By virtue of our above theorem,
we will get the same algebra (modulo gauge transformations) on the gauge theory
side (for explicit calculations see also \cite{Wolf:2004hp}).
Note that such deformations can be used for the explicit construction of 
solutions to the field equations \cite{Popov:2005uv}.

Another example is concerned with affine extensions of superconformal
symmetries. In \cite{Wolf:2004hp}, it was shown that the generators of
the superconformal algebra when viewed as vector fields have to be 
pulled back to the correspondence space (and hence to the supertwistor
space) in a very particular fashion. Their pull-backs are explicitly given by
\eqna{scgenlifted}{
           \widetilde{P}_{\a\ad}\ &=\ P_{\a\ad},\qquad
           \widetilde{Q}_{i\a}\ =\  Q_{i\a},
           \qquad
           \widetilde{Q}^i_\ad\ =\ Q^i_\ad,\\
           \widetilde{D}\ &=\ D,\\
           \widetilde{K}^{\a\ad}\ &=\ K^{\a\ad}+x^{\a\bd}Z_\bd^{~\ad},\qquad
           \widetilde{K}^{i\a}\ =\ K^{i\a},\qquad
          \widetilde{K}^\ad_i\ =\ K^\ad_i+\eta^\bd_i Z_\bd^{~\ad},\\
           \widetilde{J}_{\a\b}\ &=\ J_{\a\b},\qquad
           \widetilde{J}_{\ad\bd}\ =\ J_{\ad\bd}-\tfrac{1}{2}Z_{\ad\bd},\\
           \widetilde{T}_i^j\ &=\ T_i^j,\qquad
           \widetilde{A}\ =\ A ,
}
where the untilded quantities, commonly denoted by $N_a$ in the sequel, 
are the usual vector field expressions for the 
superconformal generators on four-dimensional superspace-time and
\eqn{scgenliftedi}{Z_{\ad\bd}\ :=\ \l^\pm_\ad\l^\pm_\bd\partial_{\l_\pm}
                  +\hat{\l}^\pm_\ad\hat{\l}^\pm_\bd\partial_{\bar{\l}_\pm}.}
Here, $(\hat{\l}^+_\ad):=\ \!^t(-\bar{\l}_+,1)$ and 
$\hl^+_\ad=\bar{\l}_-^{-1}\hl^-_\ad$. Let us define the
following (holomorphic) action on the transition function:
\eqn{}{\d^m_a f_{+-}\ :=\ \l_+^m\widetilde{N}_af_{+-},
      \qquad{\rm for}\qquad m\in\IZ,}
where $\widetilde{N}_a$ represents any of the generators given above. Note that
the $\bar{\l}$-derivative drops out as $f_{+-}$ is holomorphic in
$\l_\pm$. For 
$m=0$, we find
\eqn{}{\d^0_a\CA_{\a\ad}\ =\ \CL_{N_a}\CA_{\a\ad}\qquad{\rm and}\qquad
       \d^0_a\CA^i_\ad\ =\ \CL_{N_a}\CA^i_\ad,}
where $\CL_{N_a}$ is the Lie derivative along $N_a$. Furthermore,
one straightforwardly deduces
\eqn{}{{[\d^m_a,\d^n_b\}}f_{+-}\ =\ 
         ({f_{ab}}^c+ng_a\d^c_a-(-)^{p_a}mg_b\d^c_a)\d^{m+n}_cf_{+-},}
what represents a centerless Kac-Moody-Virasoro type algebra.
Here, the ${f_{ab}}^c$s are the structure constants of the superconformal
algebra.
The $g_a$s are abbreviations for $\l_+^{-1}\widetilde{N}^{\l_+}_a$, where 
$\widetilde{N}^{\l_+}_a$ represents the $\partial_{\l_+}$-component
of $\widetilde{N}_a$. 
Hence, this time we obtain structure functions rather than structure constants
and therefore we have to 
restrict our discussion to a certain subalgebra of the superconformal algebra 
in order to apply the above theorem. The most naive way of doing this is simply
by dropping the special conformal generators $\widetilde{K}^{\a\ad}$ and
$\widetilde{K}^\ad_i$ and the rotation generators $J_{\ad\bd}$. Then one 
eventually obtains honest structure constants and can therefore use the 
theorem. However, in \cite{Wolf:2004hp} we have seen that one need not to 
exclude the rotation generators $J_{\ad\bd}$, since the structure functions
for the maximal subalgebra of the superconformal algebra which does not contain
$\widetilde{K}^{\a\ad}$ and $\widetilde{K}^\ad_i$ are only dependent on 
$\l_\pm$. By inspecting the formulas \eqref{DeformA1}, we see that such 
$\l$-dependent functions do not spoil the generic form of the transformations
on the gauge theory side, i.e., the corresponding symmetry algebra still
closes. This is not the case when $\widetilde{K}^{\a\ad}$ and 
$\widetilde{K}^\ad_i$ are included as the structure functions also depend
on $x^{\a\ad}$ and $\eta^\ad_i$, respectively. 
For more details see \cite{Wolf:2004hp}.

Finally, let us stress that the existence of such such algebras originates 
from 
the fact that the full group of continuous transformations acting on the space 
of holomorphic vector bundles over the supertwistor space is a semi-direct 
product of the group of local 
holomorphic automorphisms (i.e., complex structure 
preserving maps) of the supertwistor space and of the group of one-cochains 
for a certain covering of the supertwistor space with values in the sheaf of 
holomorphic maps of the supertwistor space into the gauge group. This can be
shown by following and generalizing the lines presented in the case of the
purely bosonic self-dual YM equations \cite{Mason:1991rf,Popov:1998}.

\end{document}